\documentstyle[aps,prl,manuscript]{revtex}
\begin{document}
\draft
\def\dfrac#1#2{{\displaystyle{#1\over#2}}}
\title{New non-Fermi liquid type behavior given by a two band system in
normal phase.}
\author{Zsolt~Gul\'acsi and Ivan~Orlik} 
\address{ 
Department of Theoretical Physics, University Debrecen, H-4010
Debrecen, Hungary }
\date{Sept., 2000}
\maketitle
\begin{abstract}
We are reporting a new non-Fermi liquid type normal phase that has a well 
defined Fermi energy, but without showing any non-regularity in the
momentum distribution function $n_k$ in the whole momentum space,
the sharp Fermi momentum concept being undefinable. The system contains a 
natural built in gap that is visible in the physical properties of the 
system at nonzero temperatures. The presence of a flat band in multi-band 
interacting Fermi systems with more than half filling is the key feature 
leading to such a ground state, which is not restricted to one spatial 
dimension and emerges in the proximity of an insulating phase.  
\end{abstract}

Our understanding of the behavior of interacting fermionic many-body 
systems is intimately connected to the concept of Fermi-liquid introduced 
by Landau many decades ago\cite{xb1}. In a normal state that preserves all
symmetry properties of the high temperature phase of a fermionic system, the
Fermi liquid behavior has been clearly observed in the normal state of $He3$ 
and simple metals\cite{xb5}. It has the meaning that in spite of the 
inter-particle interactions, the low energy behavior of the system can be well
described within a picture of weakly interacting quasi-particles\cite{xb2}. 
This picture can be also mathematically formulated\cite{xb3}. In these terms, 
in a normal Fermi liquid: a) There is a one-to-one correspondence between the 
non-interacting one particle states and interacting single-particle states.
This is concretely obtained by describing the interacting system using a 
perturbation theory that is convergent up to infinite order. b) The single 
particle Green-functions have a quasi-particle pole that gives rise to a 
discontinuity of the momentum distribution function $n_k$ at the Fermi 
surface whose position is specified by a sharp Fermi momentum value 
$\vec{k}_F$. c) The residual quasi-particle interactions can be described by 
a small number of parameters, called Landau parameters, which can be deduced 
from a microscopic theory taking into account non-divergent two-particle 
vertex functions\cite{xb4}. In the last decade however non-Fermi liquid 
behavior has been observed experimentally in the normal phase of a variety
of materials, including higher than one dimensional systems of large interest.
Examples are: high temperature superconductors\cite{bev3}, 
heavy-fermions\cite{bev4}, layered systems\cite{bev4a}, quasi-one dimensional
conductors, doped semiconductors, systems with impurities, materials 
presenting proximity to metal-insulator transition\cite{xb2}, etc. These 
results are often discussed in terms of multi-band models\cite{bev5}, 
the presence of a some kind of gap in the normal phase 
being clearly established in many cases and subject of intensive 
studies\cite{bev5,bev6}. The last decade witnessed a huge intellectual effort
\cite{bev1} for the understanding of the non-Fermi liquid behavior in the 
normal phase\cite{bev2} of fermionic systems. On the theoretical side however,
for pure systems, the existence of a non-Fermi liquid in a normal phase has 
been exactly proved only for the one dimensional case (i.e. Luttinger 
liquid\cite{xb6}). So far, the possibility of extending the proof to two 
spatial dimensions has not been demonstrated rigorously. 
In fact, a rigorous theory of a non-Fermi liquid normal state in higher than 
one spatial dimensions is missing. For this reason, the theoretical 
understanding of different phenomena observed in the materials listed above 
is relatively poor and theoretical advance in this subject is badly
needed. Our work on the periodic Anderson model (PAM) at nonzero and finite 
on-site Coulomb repulsion (U) (a prototype of two-band systems containing 
strong correlation effects) was motivated by this state of affairs. We are
presenting in this Letter for the first time, an exact solution for this
model. The solution present in a restricted (but continuous and infinite 
domain of the phase diagram) represents a new type of non-Fermi liquid 
behavior in a normal phase for a system that has a built in gap. The obtained 
ground state energy cannot be expressed as a sum of contributions of the
on-site Hamiltonian terms, and the ground state expectation value of the 
kinetic energy terms is nonzero and negative\cite{xb6x}. The state emerges
in the vicinity of a Mott insulating phase in a continuous domain of
concentration above $3/4$ filling and has a well
defined Fermi energy, but the Fermi momentum cannot be defined, $n_k$ being 
without any non-regularity in momentum space. The property is due to the
emergence of a flat band in a multi-band system with more than half filling 
and can be extended in an exact manner to two spatial dimensions\cite{xb77}. 
Such features have been observed experimentally, e.g., in ARPES data, which,
even for high-$T_c$ materials often reflect main bands without any sharp 
characteristics in $n_k$ \cite{bev7} or necessitate the assumption of 
the presence of flat bands\cite{bev8}. Band structure calculations for 
layered systems often show a Fermi level positioned exactly at the bottom of 
a conduction band with a relatively large effective mass around its minimum, 
below which a gap is present\cite{bev4a}. Connections between the emergence 
of superconductivity and flat dispersions were also clearly pointed out in 
Ref.\cite{bev9}. Flat-band features are present in heavy-fermion systems as 
well \cite{bev10}, and can be even produced by squashing carbon nanotubes
\cite{bev11}. 
 
We consider two bands denoted by $b \: = \: c , \: f$, the starting 1D
Hamiltonian being $\hat{H} \: = \: \hat{H}_0 \: + \: \hat{U}$. The Hubbard 
term is $\hat{U} \: = \: U \sum_i \: \hat{n}^f_{i,\uparrow} \: 
\hat{n}^f_{i,\downarrow}$ and we have $\hat{H}_0 \: = \: \hat{T}_c \: 
+ \: \hat{T}_f \: + \: \hat{E}_f \: + \: \hat{H}_{h}$
with the kinetic energies $\hat{T}_b \: = \: t_b \sum_{i,\sigma} \: \left[ \: 
b^{\dagger}_{i,\sigma} b_{i+1,\sigma} \: + \: h.c. \: \right]$, 
the on-site $f$ level energy $\hat{E}_f \: = \: E_f \: \sum_{i,\sigma} 
\: \hat{n}_{i,\sigma}^f$, hybridization energy 
$\hat{H}_h \: = \: \hat{V}_0 \: + \: \hat{V}_1$, and
\begin{eqnarray}
\hat{V}_0 \: & \: = \: & \: \sum_{i,\sigma} \: \left[ \: \tilde{V}_0 \:
c^{\dagger}_{i,\sigma} \: f_{i,\sigma} \: + \: h.c. \: \right] \: , 
\nonumber\\
\hat{V}_1 \: & = & \: \sum_{i,\sigma} \: \left[ \: \tilde{V}_1 \: \left( \:
c^{\dagger}_{i,\sigma} \: f_{i+1,\sigma} \: + \: f^{\dagger}_{i,\sigma} \:
c_{i+1,\sigma} \: \right) \: + \: h.c. \: \right] \: .
\label{x1}
\end{eqnarray}
We are presenting the case when $\hat{H}_h$ contains 
imaginary coupling constants, i.e. 
$\tilde{V}_0 \: = \: i \: V_0$ and $\tilde{V}_1 \: = \: i \: V_1$, where 
$V_0, \: V_1$ are real. For the starting point we consider $3/4$ filling. In 
this Letter we shall give an exact solution of this problem, valid in a 
restricted domain of the phase diagram relevant for our study. 
In order to have a clear image of the solution and its physical meaning, let 
us consider the $U \: = \: 0$ case first. To this
end, we would like to express $\hat{H}_0$ through a $\hat B \: = \: 
\sum_{i,\sigma} \: B^{\dagger}_{i,\sigma} \: B_{i,\sigma}$ term, where
$B_{i,\sigma} \: = \: \alpha \: c_{i,\sigma} \: + \: \beta \: c_{i+1,\sigma} 
\: + \: \gamma \: f_{i,\sigma} \: + \: \delta \: f_{i+1,\sigma}$
and $\alpha, \: \beta, \: \gamma, \: \delta$ are constants to be determined.
Introducing the total particle number operator $\hat{N} \: = \: 
\sum_{i, \sigma, b} \: \hat{n}^b_{i,\sigma} $, we find then that 
$ \hat{H}_0 \: = \: - \: \hat B \: + \: \eta \: \hat{N} $ 
holds, if the following relations are satisfied
\begin{eqnarray}
\eta \: & \: = \: & \: |\alpha|^2 \: + \: |\beta|^2 \: ,
\label{x4}
\\
E_f \: & \: = \: & \: \eta \: - \: \left( \: |\gamma|^2 \: + \: 
|\delta|^2 \: \right) \: ,
\label{x5}
\\
- \: t_c \: & \: = \: & \: \alpha^* \: \beta \: = \: \beta^* \: \alpha \: , 
\label{x6}
\\
- \: t_f \: & \: = \: & \: \gamma^* \: \delta \: = \: \delta^* \: \gamma \: , 
\label{x7}
\\
i \: V_0 \: & \: = \: & \: \alpha \: \gamma^* \: + \: \delta^* \: \beta \: = 
\: - \: ( \: \alpha^* \: \gamma \: + \: \delta \: \beta^* \: ) \: , 
\label{x8}
\\
i \: V_1 \: & \: = \: & \: \alpha \: \delta^* \: = \: \gamma \: \beta^* \: = 
\: - \: \alpha^* \: \delta \: = \: - \: \gamma^* \: \beta \: . 
\label{x9}
\end{eqnarray}
Eqs.(\ref{x6} - \ref{x9}) determine through  $\alpha, \: \beta,
\: \gamma, \: \delta$ the value of $E_f$ and $\eta$ for which the presented
structure of $\hat{H}_0$ is valid. With 
$\eta \: = \: 2 \: \sqrt{ \bar{v}^2 \: + \: t_c^2 }, \: $
$\: \bar{v} \: = \: V_0 \: V_1 \: / \: ( \: 2 \: t_f )$ 
we obtain 
\begin{eqnarray}
E_f \: = \: \eta \: ( \: 1 \: - \: t_f^2 \: V_1^{-2} \: ) \: , \quad 
V_1^2 \: = \: - \: t_c \: t_f \: .
\label{x10}
\end{eqnarray}
Taking now into consideration that
$B^{\dagger}_{i,\sigma} \: B_{i,\sigma} \: + \:
B_{i,\sigma} \: B^{\dagger}_{i,\sigma} \: = \: |\alpha|^2 \: + \: |\beta|^2 
\: + \: |\gamma|^2 \: + \: |\delta|^2$ and imposing the condition 
$\langle \: \hat{N} \: \rangle \: = \: 3 \: L$ where $L$ 
represents the number of lattice sites ($3/4$ filling), we find
$\hat{H}_0 \: = \: E_g \: + \hat P \: ,$ 
where $E_g \: = \: L \: \eta \: \left( \: 
1 \: - \: 2 \: |m|^2 \: \right)$ and  
$m \: = \: t_f \: / \: \tilde{V}_1$ has been introduced. The operator 
$\hat P \: = \: \sum_{i,\sigma} \: B_{i,\sigma} \: B^{\dagger}_{i,\sigma}$ 
being positive semidefinite, $E_g$ is the ground state energy, and the
ground state wave function is that $| \: \psi_g \: \rangle$, for which we 
have $\hat P \: | \: \psi_g \: \rangle \: = \: 0$. We demonstrate now that 
\begin{eqnarray}
| \: \psi_g \: \rangle \: = \: \prod_{i = 1}^{L} \: \left( \: 
\prod_{\alpha =1}^3 F^{(\alpha)}_i \: \right) \: | \: 0 \: \rangle \: ,
\label{x12}
\end{eqnarray}
where $| \: 0 \: \rangle$ is the bare vacuum with no fermions present,
$ \: D_{i,\sigma} \: = \: \alpha^* \: \left( \: c^{\dagger}_{i,\sigma} \: + 
\: m  \: f^{\dagger}_{i,\sigma} \: \right) \: + \: \beta^* \: \left( \: 
c^{\dagger}_{i+1,\sigma} \: + \: m^* \: f^{\dagger}_{i+1,\sigma} \: \right)$,
and $F^{(\alpha=1,2)}_i \: = \: D_{i,\sigma(\alpha)=\uparrow , \downarrow}.$
In order to see this, one can easily verify that
$B^{\dagger}_{j,\sigma} \: F_{i}^{(\alpha)} \: = \: - \: 
F_{i}^{(\alpha)} \: B^{\dagger}_{j,\sigma}$ independent of the indices, 
and $B^{\dagger}_{i,\sigma} \: D_{i,\sigma} \: = \: 0$, so $ \hat P \: | \: 
\psi_g \: \rangle \: = \: 0$, and $| \: \psi_g \: \rangle$ is the ground
state. We must stress, that for $U \: = \: 0$ the ground state is entirely
given by $F^{(1)}_i$ and $F^{(2)}_i$, the $F^{(3)}_i$ operator being 
completely arbitrary apart from the requirement that it introduces
under the $\prod_i$ product $L$ electrons into the system. The concrete
expression of $F^{(3)}_i$ is fixed by the nonzero $U$ as follows. When 
$U \: \ne \: 0$, the Hamiltonian contains, besides $\hat{H}_0$, 
the Hubbard $\hat{U}$ as well. However, one may observe that $\hat{U}$ can be 
exactly transformed as
$
\hat{U} \: = \:  U \: \hat{P}' \: + \: 
U \: \sum_{i,\sigma} \: \hat{n}^f_{i,\sigma} \: - \: U \: L \: ,
$
where $\hat{P}'= \sum_i \: \hat{P}'_i$ and
$\hat{P}'_i \: = \: ( \: 1 \: - \: \hat{n}^f_{i,\uparrow} \: - \: 
\hat{n}^f_{i,\downarrow} \: +  \: \hat{n}^f_{i,\uparrow} \: 
\hat{n}^f_{i,\downarrow})$. But, $\hat{P}'_i$ is one if on the $i$ site 
there are no $f$ electrons, and is zero, if on site $i$ there is at least one 
$f$ electron. As a consequence, $\hat{P}'$ gives a 
sum of non-negative numbers, so it is a positive semidefinite operator. 
Furthermore, $\hat{P}'$ gives its minimum eigenvalue (i.e. zero) for a wave 
function that contains at least one $f$ electron on every site of the 
lattice. Let us consider for this reason
\begin{eqnarray} 
F^{(3)}_i \: = \: \sum_{\sigma} \: a_{\sigma} \: f^{\dagger}_{i,\sigma} \: ,
\label{xxxxx}
\end{eqnarray}
where the $a_{\sigma}$ are constants to be determined.
Now $| \: \psi_g \: \rangle$ introduces 3 electrons $L$ times within the 
system, so the solution (two bands are present) is indeed for $3/4$ filling. 
Because of Eq.(\ref{xxxxx}) the 3 electrons per lattice site are distributed 
in the ground state in such a way, that on every site we have always at least 
one $f$ electron present.  As a consequence,
$\hat{H} \: = \: \hat{H}_0 \: + \: \hat{U} \: = \: [ \: \hat{H}_0 \: + \:
U \: \sum_{i,\sigma} \: \hat{n}^f_{i,\sigma} \: - \: U \: L \: ] \: 
+ \: U \: \hat{P}'$ has the ground state wave function given by 
Eqs.(\ref{x12},\ref{xxxxx}), provided the energy of the $f$ 
level in $\hat{H}_0$ is renormalized as $E_f' \: = \: E_f \: + \: U$ and 
$E_g$ is shifted down with $U \: L$. We underline that in the $U \: 
\to \: 0$ limit, the described $| \: \psi_g \: \rangle$ becomes only a small 
contribution from the linear combination of wave functions that build up the 
ground state. The physical properties of the interacting ground state are 
present only at $U > 0$, and the $U \: \ne \: 0$ state cannot be 
perturbatively obtained from the $U \: = \: 0$ case. This confirms the general
belief that a non-Fermi liquid emergence has to be a clear non-perturbative 
effect \cite{fuk}.

For $U \: > \: 0$, instead of $\hat H_0$ we have
$
\hat{H} \: = \: - \: \hat B \: + \: \eta \: \hat{N} \: - 
\: U \: L \: + \: U \: \hat{P}'
$
and Eqs.(\ref{x4} -\ref{x9}) remain all valid, excepting Eq.(\ref{x5}),
which becomes $E_f' \: = \: E_f \: + \: U \: = \: |\alpha|^2 \:
+ \: |\beta|^2 \: - \: |\gamma|^2 \: - \: |\delta|^2$. Because 
Eqs.(\ref{x6}-\ref{x9}) remain the same, the $\alpha, \: \beta, \: \gamma,
\: \delta, \: \eta, \: m$ values remain unaltered (together with $e_1, \: e_2$
in Eq.(\ref{x24}) below),  we obtain
$\hat{H}'_0 \: = \: E_g^U \: + \: \hat{P}_{+} \: $, where
$E_g^U \: = \: L \: \eta \: \left( \: 1 \: 
- \: 2 \: |m|^2 \: \right) \: - \: U \: L $. 
The term $\hat{P}_{+} \: = \: \hat P \: + \: U \: \hat{P}'$ is a positive 
semidefinite operator, whose minimum eigenvalue is given by 
$| \: \psi_g \: \rangle$ since $\hat{P}_{+} \: | \: \psi_g \: \rangle \: =
\: 0$. The conditions that enable the construction of the solution for 
$\hat{H}$ change from Eq.(\ref{x10}) to 
\begin{eqnarray}
E_f' \: = \: E_f \: + \: U \: = \: \eta \: ( \: 1 \: - \: |m|^2  \: ) \: , 
\label{x36}
\end{eqnarray}
$\eta$ and $V_1^2$ remaining as given in Eq.(\ref{x10}) and $U \: > \: 0$.
We denote the manifold defined by Eq.(\ref{x36}) by $D_p$ \cite{ex1}.

In fact $| \: \psi_g \: \rangle$ can be written in $k$ space as well. 
Fourier transforming and taking $e_1 = \: \alpha^* \: + \: 
\beta^* \: \exp{(ik)}$, $e_2 = \: \alpha^* \: m \: + \: \beta^* \: m^* \:
\exp{(ik)}$, $F^{(\alpha)}_i = \sum_{k} \: e^{ i k r_i} \: 
F_{k}^{(\alpha)}$,
$D_{k,\sigma} 
\: = \: e_1 \: c^{\dagger}_{k,\sigma} \: + \: e_2 \: f^{\dagger}_{k,\sigma}, $
$\: F_k^{(3)} \: = \: \sum_{\sigma} \: a_{\sigma} \: 
f^{\dagger}_{k,\sigma}$,
with $F_{k}^{(\alpha)} \: F_{k}^{(\alpha)} \: = \: 0$, 
$F_{k}^{(\alpha)} \: F_{k'}^{(\alpha')} \: =
\: - \: F_{k'}^{(\alpha')} \: F_{k}^{(\alpha)}$, and
$M \: = \: \sum_P \: (-1)^p \: exp [ \: i \: ( \: r_1 \: k_{i_1} \: + 
\: r_2 \: k_{i_2} \: + ... + \: r_L \: k_{i_L} \: ) \: ]$ where $\sum_P$ 
denotes a sum over all permutations of $(1, \: 2, ... \: L)$ to 
$(i_1, \: i_2,... \: i_L)$ and $p$ represents the number of pair-permutations
in a given $P$, we find
$ | \: \psi_g \: \rangle \: = \: M^3 \: \prod_k \: \left( \: 
\prod_{\alpha=1}^3 \: F_{k}^{(\alpha)} \: \right) \: | \: 0 \: \rangle \: $ .
With $A=M^3$ we finally obtain
$| \: \psi_g \: \rangle \: = \: A \: \prod_k \: d_k \: | \: 0 \: \rangle$,
where
\begin{eqnarray}
&& d_k \: = \: e_1^2 \: c^{\dagger}_{k,\uparrow} \: 
c^{\dagger}_{k,\downarrow} \: F^{(3)}_k  \: + \:
e_1 \: e_2 \: f^{\dagger}_{k,\downarrow}
\: f^{\dagger}_{k,\uparrow} \: \sum_{\sigma} \: a_{\sigma} \:
c^{\dagger}_{k,\sigma} \: ,
\label{x24}
\end{eqnarray}
and
$
\langle \: \psi_g \: | \: \psi_g \: \rangle \: = \: |A|^2 \: \prod_k \: 
|e_3|^2 \: \left[ \: |e_1|^4 \: + \: |e_1|^2 \: |e_2|^2 \: \right] \: .
$
The $e_3$ constant coefficient is chosen to preserve the 
normalization to unity \cite{ex2} and 
$|e_3|^2 \: = \: \sum_{\sigma} |a_{\sigma}|^2 $. 
  
To have an insight about the physical behavior of the system in the ground 
state, let us diagonalize first $\hat{H}_0$ in momentum space. Denoting by
$H_k^h \: = \: ( \: V_k \: c^{\dagger}_{k,\sigma} \: f_{k,\sigma} \: + \:
h.c. \: )$, in the thermodynamic limit we find
\begin{eqnarray}
\frac{\hat{H}_0}{L} \: = \: \sum_{\sigma} \: \int_{0}^{2\pi} \:
\frac{dk}{2\pi} \: \left[ \: \sum_{b=c,f} \: \epsilon^b_k \: 
b^{\dagger}_{k,\sigma} \: b_{k,\sigma} \: + \:  H_k^h \: \right] \: ,
\label{x27}
\end{eqnarray}
where 
$\epsilon^c_k \: = \: 2 \: t_c \: \cos{k}, \; \epsilon^f_k \: = \: E_f \: + \:
2 \: t_f \: \cos{k}, \; V_k \: = \: 2 \: V_1 \: \sin{k} \: + i \: V_0$. 
Introducing the row vector $W^{\dagger}_k \: = \: ( \: c^{\dagger}_{k,\sigma},
\: f^{\dagger}_{k,\sigma} \: )$ and the $(\: 2 x 2 \: )$ matrix
$\tilde{R}$ with components $R_{(1,1)}\: = \: \epsilon^c_k, \; 
R_{(1,2)} \: = \: V_k, \; R_{(2,1)} \: = \: V^*_k, \; R_{(2,2)} \: 
= \: \epsilon^f_k$, we may write the integrand in 
Eq.(\ref{x27}) as $W^{\dagger}_k \: \tilde{R} \: W_k$. The diagonalization
in $k$ space reduces to the secular equation written for $\tilde{R}$. 
Two bands arise, that via Eq.(\ref{x10}) becomes
\begin{eqnarray}
E^{(1)}_k \: & = & \: 2 \: \sqrt{ \bar{v}^2 \: + \: t^2_c} \: > \: 0 \: ,
\nonumber\\
E^{(2)}_k \: & = & \: 2 \: \frac{t_f}{t_c}\: \sqrt{ \bar{v}^2 \: + \: t^2_c} 
\: + \: 2 \: ( \: t_f \: + \: t_c \: ) \: \cos{k} \: ,
\label{x31}
\end{eqnarray}
where we have as presented in Eq.(\ref{x10}), $sign(t_f) \: = \: - sign(t_c)$.
The system is at $3/4$ filling, so the lower band $E^{(2)}_k$ is completely
filled, and the upper band $E^{(1)}_k$, which is completely flat, is half
filled. There is no hybridization between these two 
bands and taking into account their filling, the ground state energy $E_g$ is
re-obtained. In the presence of $U \: > \: 0$, in the ground state, because 
$\hat{P}' \: | \: \psi_g \: \rangle \: = \: 0$, 
the effective Hamiltonian is in fact $\hat{H}_0'/ L$ which differs from 
$\hat{H}_0 / L$ in that it has a renormalized $E_f'=E_f + U$,
and its energy scale is shifted down with $U$. Effectuating the calculations
for the band structure as presented for Eq.(\ref{x31}) but using instead of 
$E_f$ the $E_f'$ value, we re-obtain (shifted down with U) for the 
ground state the structure presented in Eq.(\ref{x31}). So we have a well 
defined Fermi energy, positioned at $E^{(1)}_k=$ constant, the Fermi momentum
being undefinable. The system has a natural built in gap (the minimum and
nonzero distance between the $E^{(1)}$ and $E^{(2)}$) that will be visible in
the physical properties at $T\ne 0$ \cite{ex3}. 

To understand the physical behavior of the system all ground state 
expectation values at 
$U \: > \: 0$ relevant for our study can be expressed from Eq.(\ref{x24}).
Introducing $Z^{-1}_k \: = \: ( \: |e_1|^4 \: + |e_1|^2 \: |e_2|^2 \: ) \:
|e_3|^2$, we have
\begin{eqnarray}
\langle \: f^{\dagger}_{k,\sigma} \: f_{k,\sigma} \: \rangle \: & = & \: 
\left[\: |a_{\sigma}|^2 \: |e_1|^4 \: + |e_3|^2 \: |e_2|^2 \: |e_1|^2 \: 
\right] Z_k \: ,    
\nonumber\\
\langle \: c^{\dagger}_{k,\sigma} \: c_{k,\sigma} \: \rangle \: & = & \:
\left[ \: |a_{\sigma}|^2 \: |e_1|^2 \: |e_2|^2 \: + \: |e_3|^2 \: |e_1|^4 
\: \right] Z_k \: ,  
\nonumber\\
\langle \: c^{\dagger}_{k,\sigma} \: f_{k,\sigma} \: \rangle \: & = & \:
\left[ \: |a_{- \sigma}|^2 \: |e_1|^2 \: e_1^* \: e_2 \: \right] Z_k .
\label{x37}
\end{eqnarray}
It is seen that $n_k \: = \: \sum_{\sigma} \: \left( \: \langle \:
c^{\dagger}_{k,\sigma} \: c_{k,\sigma} \: \rangle \: + \: \langle \:
f^{\dagger}_{k,\sigma} \: f_{k,\sigma} \: \rangle \: \right) \: = \: 3$
i.e. the total momentum distribution function is uniform in $k$ space, 
so the system is a non-Fermi liquid. 
From Eq.(\ref{x37}) all individual contributions in $n_k$ can be expressed.
We obtain functions of $k$ that are continuous together with 
their derivatives of any order in the whole momentum space. For example,
$
n^c_{k,\uparrow} \: = \: \langle \: c^{\dagger}_{k,\uparrow} \: 
c_{k,\uparrow} \: \rangle \: = \: 1 \: - \: A \: ( \eta \: + \: 2 \:
t_c \: \cos{k} ) (\bar{\eta} \: + \: 2 \: t_c \: \cos{k} )^{-1} \: , 
$
where the constants are $A \: = \: |a_{\downarrow}|^2 \: |m|^2 \: / \: 
[ |e_3|^2 \: (|m|^2 \: - \: 1 ) ]$ and $\bar{\eta} \: = \: \eta \: 
( |m|^2 \: + \: 1 ) / ( |m|^2 \: - \: 1 ) \: > \: 2 t_c$. The movement of 
particles is allowed by $| \: \psi_g \: \rangle$, that requires 
,,at least one $f$ electron on every site'', i.e. allows nonzero hopping 
matrix elements. Indeed, starting from Eq.(\ref{x37}) the ground state 
expectation values of all contributions to $\hat{H}$ can be expressed. With 
$I \: / \: L \: = \: \int_0^{2\pi} \: dk \: [ \: 2 \: \pi \: ( \: |e_1|^2 
\: + \: |e_2|^2 \: ) \: ]^{-1} > \: 0 $, we find
\begin{eqnarray}
\frac{ \langle \: \hat{T}_c \: \rangle }{ A_1 } \: & = & \: 2 \: \eta \: 
|m|^2 , \;
\frac{ \langle \: \hat{T}_f \: \rangle }{ A_1 } \: = \: 2 \: \eta \: |m|^4 , 
\;
\frac{ \langle \: \hat{E}_f \: \rangle }{ A_2} \: = \: E_f \: ,
\nonumber\\
\langle \: \hat{U} \: \rangle \: & = & \: U \: \left( \: A_2 \: - \: 
L \: \right) , \quad
\langle \: \hat{V}_0 \: \rangle \: = \: - \: 2 \: V_0^2 \: I \: ,
\nonumber\\
\langle \: \hat{V}_1 \: \rangle \: & = & \: - \: \frac{ 2 \: \eta \: |m|^2 \:
( \: 1 \: + \: |m|^2 \: ) \: L }{ ( \: 1 \: - \: |m|^2 )^2 } \: + \: A_3 \: ,
\label{x40}
\end{eqnarray}
where  
$A_1 \: = \: [ \: L \: - \: \eta \: ( \: 1 \: + \: |m|^2 \: ) \: I \: ] \: 
/ \: ( \: 1 \: - \: |m|^2 )^2$, $A_2 \: = \: [ \: ( \: 1 \: - \: 2 \: |m|^2
\: ) \: L \: + \: 2 \: \eta \: |m|^2 \: I \: ] \: / \: ( \: 1 \: - \: |m|^2)$,
and
$A_3 \: = \: \{ \: 2 \: \eta^2 \: [ \: ( \: 1 \: + \: |m|^2 \: ) \: / \: 
( \: 1 \: - \: |m|^2 \: ) \: ]^2 \: - \: 8 \: V_1^2 \: \} \: I $.
Summing up all contributions in Eq.(\ref{x40}) we re-obtain exactly the ground
state energy $E_g^U$. We observe from Eq.(\ref{x40}) that 
$E_g^U$ cannot be expressed as a sum of on-site contributions. With
$|\bar{m}|^2 \: = \: |m|^2 \: ( \: 1 \: + \: |m|^2 \: ) \: / \: 
( \: 1 \: - \: |m|^2 \: )^2, $ $ \: J \: = \: \eta \: ( \: 1 \: + \: |m|^2 \:
) \: I \: - \: L, $ 
$\langle \hat H_{loc} \rangle \: = \:
\langle \: \hat{E}_f \: \rangle \: + \: \langle \: \hat{U} \: \rangle \: + 
\: \langle \: \hat{V}_0 \: \rangle, $
we have
\begin{eqnarray}
&& \langle \: \hat{E}_f \: \rangle \: + \: \langle \: \hat{U} \: \rangle \: + 
\: \langle \: \hat{H}_h \: \rangle \: = \: E_g^U \: + 
2 \: \eta \: |\bar{m}|^2 \: J \: > \: E_g^U \: ,
\nonumber\\
&& \langle \hat H_{loc} \rangle \:
 = \: E_g^U \: + \: 8 \: V_1^2 \: I \:
> \: E_g^U \: . 
\label{x41} 
\end{eqnarray}
Here $J \: = L \: [ \: 1 \: - \: r^{-2} \: ]^{-1/2} - L \: > \: 0$ 
and $r \: = ( \: \eta \: / \: 2 \: t_c \:
) \: [ \: ( \: |m|^2 \: + \: 1 \: ) / ( \: |m|^2 \: - \: 1 \: ) \: ]$ .
From Eq.(\ref{x41}) it can be seen that the system is not localized
because $ \langle \hat H_{loc} \rangle > E_g^U$, i.e. the sum of the 
ground-state expectation values of all on-site (localized) terms from the 
Hamiltonian is greater than the ground-state energy itself.
On the other hand, from Eq.(\ref{x40}) 
$
\sum_{b=c,f} \: \langle \: \hat{T}_b \: \rangle \: = \: - \: 2 \: \eta \: 
|\bar{m}|^2 \: J \: < \: 0 \: .
$
Furthermore, as can be seen from Eq.(\ref{x40}, \ref{x41}) adding the 
non-local part of the hybridization to the kinetic energy contributions
$\langle \hat T_c \rangle + \langle \hat T_f \rangle$
we obtain also a negative number
$\langle \hat H_{mov} \rangle = \langle \hat{T}_c \rangle + 
\langle \hat{T}_f \rangle +
\langle \hat V_1 \rangle < 0$, \cite{Gu1}. So taken together, the sum of 
all ground-state
expectation values connected to the movement of particles within the
system is negative. As a consequence, similar to the 2D case with flat upper
band and imaginary $m$ parameter \cite{Gu}, in the situation described here,
the system is maintaining its itinerant character instead of becoming 
insulating, because the $\langle \hat H_{mov} \rangle$ (i.e. the movement of
particles within the system) decreases the total energy.
As a consequence, the system is in a normal phase, but without having any 
non-regularity in $n_k$ at any point in $k$ space (even in its derivatives 
with respect to $k$). From Eq.(\ref{xxxxx}), the coefficient $a_{\uparrow}$ 
remains undetermined, the ground state has a large spin degeneracy, so it
is paramagnetic (see also Ref. \cite{ex2}). 

From the point of view of the described properties is important to analyze 
the system also around $D_p$. Concerning these aspects,
first of all it can be seen that the presented state emerges also in the
presence of doping. Taking $N = 3L + n_r$, where $1 \leq n_r < L$ represents 
the additional electrons introduced into the system above $3/4$ filling,
we define $\hat F^{(4)} = \sum_{\{k\}} A_{\{k\}} \prod_{k}^{n_r} f^{(4)}_k$ 
with 
$
f^{(4)}_k = \epsilon_{\uparrow} (f^+_{k,\uparrow} + 
e^{i\phi_c} c^+_{k,\uparrow}) + \epsilon_{\downarrow} e^{i\phi_{\downarrow}}
(f^+_{k,\downarrow} + e^{i\phi_c} c^+_{k,\downarrow}) \: ,
$
where $A_{\{k\}}$ represents the coefficient of a given $\{k\}$ combination 
of $n_r$ different and ordered $k_i$ taken from the $L$ possible $k$-values.
The ground state wave function becomes under doping
$|\psi_{gd} \rangle = 
\prod_i(\prod_{\alpha=1}^3 F^{(\alpha)}_i) F^{(4)} |0 \rangle$, the 
character of the phase described remaining unchanged since the first three
operatorial components from $|\psi_{gd} \rangle$ remain as in Eq.(\ref{x12}).
As can be seen, the state we are describing persists also above half-filling
concentration of particles within the upper band.
Concerning the $N$ dependence, we further mention that the system is a
Fermi-liquid for $N < 2L$. 

Taking into account that
the flat-band feature is intimately connected only to the presence of the
$\hat B$ term in the Hamiltonian ($\hat B$ collecting all $k$ dependences), 
we can maintain the described flat-band features together with the non-Fermi 
liquid properties taking into consideration some local or global contributions
to $\hat H$ that keep the system itinerant (otherwise the system becomes
insulating). For example, we can 
leave the $D_p$ manifold taking into account arbitrary and 
independent deviations $\delta E_f$ and $\delta U$ from the $E_f$ and $U$ 
values that satisfies Eq.(\ref{x36}). Denoting by $\delta X = \delta E_f +
\delta U$, considering $U'= U + \delta U > 0$, the new Hamiltonian becomes
$\hat H'= \hat H + \delta X \sum_{i,\sigma} \hat n_{i,\sigma}^f $. Adding and 
subtracting $\delta X \hat N_c$, a new positive semidefinite operator
emerges (we consider here $\delta X > 0$) 
$\hat P_{X} = \delta X \sum_{i,\sigma}(1-c^{+}_{i,\sigma} c_{i,\sigma})$,
and we have $\hat H' = (\hat P_{+} + \hat P_{X}) + E_g'$, where taking also 
doping into account,
$E_g'/L = \eta[1-2|m|^2+(n_r/L)(1+\delta X/\eta)]-U'$. The new ground state 
wave function is obtained as $P_c |\psi_{gd}' \rangle$, where in 
$|\psi_{gd}'\rangle$ the $F^{(4)}$ term introduces only $f$-electrons, and
$\hat P_c = \prod_{i,\sigma} \hat n_{i,\sigma}^c$ maximize the number of 
$c$-electrons \cite{ex4}. Under these conditions, at $n_r = 0$,
the obtained ground state wave function is
a Mott insulator since it sets rigorously three electrons on every site of the 
lattice (similar result is obtained at $\delta X = n_r = 0$ and real 
hybridization coupling constants). As a consequence, 
the described non-Fermi liquid state emerges in parameter space in the  
close vicinity of an insulating phase. In the presence of doping, at 
$\delta X > 0$ the on-site expectation value
$\langle \hat H_{loc} \rangle = \: ( \eta - |m|^2 + \delta X )
( L + n_r) - U'L > E_g'$, the system is itinerant, and since the flat-band
feature is present, it is again non-Fermi liquid.

To further emphasize the itinerant character of the ground-state described
here we are presenting a localized ground-state obtained at $3/4$ filling
in another region of the phase diagram, but the same model, corresponding also
to a completely flat upper band \cite{Gu2}

\begin{eqnarray}
| \psi_1 \rangle = \prod_i [
(\hat c^{\dagger}_{i,\uparrow} + m \hat f^{\dagger}_{i,\uparrow})
(\hat c^{\dagger}_{i,\downarrow} + m \hat f^{\dagger}_{i,\downarrow})
(\alpha_i \hat c^{\dagger}_{i,\uparrow} + 
\beta_i \hat c^{\dagger}_{i,\downarrow} +
\gamma_i \hat f^{\dagger}_{i,\uparrow} + 
\delta_i \hat f^{\dagger}_{i,\downarrow}) ] | 0 \rangle \: ,
\label{gul1}
\end{eqnarray} 
where $\alpha_i,\beta_i,\gamma_i,\delta_i$ are constant numbers.
This ground-state contains the same number of particles (i.e. $3$) on every 
lattice site and gives \cite{Gu2}
\begin{eqnarray}
\langle \hat T_c \rangle & = & \langle \hat T_f \rangle = \langle \hat V_1 \rangle
= 0 \: , 
\label{gul21}
\\
\langle \hat H_{loc} \rangle & = & E_{g,l}^U \: ,
\label{gul22}
\end{eqnarray}
where, for the localized case, the ground-state energy per site \cite{Gu2} is
$E_{g,l}^U/L = - (V_o V_1/t_f)(1-2t_f^2/V_1^2) - U$. A comparison of the 
two ground-states from Eq.(\ref{x12}) and Eq.(\ref{gul1}) shows that
$| \psi_1 \rangle$ sets rigorously three electrons on each site of the 
lattice, while $| \psi_g \rangle$, besides sites with three electrons,
contains sites with four and two electrons as well. As a consequence, any 
hopping 
like Hamiltonian term (as $\hat T_c$, $\hat T_f$ or the ,,band-change hopping''
$\hat V_1$) applied to $|\psi_1\rangle$ gives a state orthogonal to 
$| \psi_1 \rangle$, from where Eq.(\ref{gul21}) arises. This is no more 
true in case of $| \psi_g \rangle$, from where the main difference 
between the two ground-states emerges.
From the point of view of the Kubo formula, we mention, that starting from it
we have (see for example \cite{v1})
$1/m^* \sim - \langle \hat H_{kin} \rangle$ and the real part of the 
conductivity in 
$\omega \to 0$ limit $\sigma'(\omega) \sim 1/m^*$, where $m^*$ is the
effective mass, and $\hat H_{kin}$ is the global kinetic energy (in our case
build up from $T_c, \: T_f$ and $V_1$ which give contribution in the current
operator). As a consequence, we have a localized state
only if $m^* = \infty$, i.e. in $|\psi_1\rangle$ case where Eq.(\ref{gul21}) 
or Eq.(\ref{gul22}) hold.

In conclusion, in this Letter a new non-Fermi liquid normal phase 
has been presented, which is not intimately connected to one spatial 
dimensions.
 
For Zs.G. research supported by OTKA-T-022874 and FKFP-0471. 
He is also grateful to P.F. de Chatel and M. Gulacsi for valuable discussions
and critical reading of the manuscript.


\begin{references}
\bibitem{xb6x}
As a consequence, the system is not insulating.
\bibitem{ex1}
$D_p$ is continuous, infinite and crosses the parameter space from the 
low $U$ to the high $U$ limits.
\bibitem{Gu1} This is because $\langle \hat H_{loc} \rangle + 
\langle \hat H_{mov} \rangle = E_g^U$, and $\langle \hat H_{loc} \rangle
> E^U_g$.
\bibitem{ex2}
In the text we considered $a=a_{\uparrow}$ a constant. However, in 
Eq.(\ref{xxxxx}) the $a_{\sigma}$ can be even local. Using the notation 
$|\psi_g(a)\rangle$ for Eq.(\ref{x24}), considering $\{a_k\}$ arbitrary 
k-dependent sets, the complete ground-state is
$|\psi\rangle=\sum_{\{a_k\}} \alpha(\{a_k\}) |\psi_g(\{a_k\})\rangle$
with $\langle\psi|\psi\rangle=1$. This reproduces  
Eq.(\ref{x40}), $E_g^U$, $n_k=3$ and main properties reported.
\bibitem{ex3}
Energy is absorbed by the lower band during excitations.
\bibitem{ex4} $|\psi_{gd}' \rangle$ is the ground state if after the 
normalization to unity the thermodynamic limit is taken. However, there are
special cases (for example: $t_c > 0$, $V_0 = 0$, $L$ odd) for which it
represents the ground-state even at finite $L$.
\end{references}
\end{document}